\documentclass[twocolumn,showpacs]{revtex4}

\usepackage{graphicx}%
\usepackage{dcolumn}
\usepackage{amsmath}

\makeatletter
\def\btt#1{\texttt{\@backslashchar#1}}%
\DeclareRobustCommand\bblash{\btt{\@backslashchar}}%
\makeatother

\begin{document}
\title{
Wave-vector dependent intensity variations of the Kondo peak in photoemission from CePd$_3$}
\author{
S. Danzenb\"acher,$^1$ Yu. Kucherenko,$^2$ M. Heber,$^{1*}$ D.V.
Vyalikh,$^1$ S.L. Molodtsov,$^1$ V.D.P.
Servedio,$^{1**}$ and C. Laubschat$^1$\\
{\it (1) Institut f\"ur Festk\"orperphysik, TU Dresden,
D-01062 Dresden, Germany} \\
{\it (2) Institute of Metal Physics, National Academy of Sciences
of Ukraine, UA-03142 Kiev, Ukraine}\\
{\it (*) present address: Freudenberg Dichtungs- und \mbox {Schwingungstechnik KG}, Technisches Entwicklungszentrum, D-69465 Weinheim, Germany}\\
{\it (**) present address: Sezione INFM and Dip.\ di Fisica,
Universit\`a ``La Sapienza'', P.le A.~Moro 2, 00185
Roma and Centro Studi e Ricerche e Museo della Fisica  E. Fermi, Compendio Viminale, Roma, Italy}\\
}
\date{\today}

\begin{abstract}
Strong angle-dependent intensity variations of the Fermi-level
feature are observed in $4d \rightarrow 4f$ resonant photoemission
spectra of CePd$_3$(111), that reveal the periodicity of the
lattice and largest intensity close to the $\overline{\Gamma}$
points of the surface Brillouin zone. In the framework of  a
simplified periodic Anderson model the phenomena may
quantitatively be described by a wave-vector dependence of the
electron hopping matrix elements caused by Fermi-level crossings
of non-$4f$-derived energy bands.
\end{abstract}

\begin{pacs}
{79.60.-i, 71.20.Eh, 71.27.+a}
\end{pacs}

\maketitle

In Ce compounds, interaction of localized 4$f$ states with itinerant valence-band (VB) states may
lead to a number of fascinating phenomena ranging from valence instabilities to heavy-fermion and
even non-Fermi-liquid behavior~\cite{AAA}. Although part of physical properties of these compounds
may be understood in the framework of local-density approximation (LDA) band-structure
calculations, the latter are not able to account properly for the correlated nature of the $f$
electrons. For the latter, localized approaches are applied based on the Kondo or Anderson
hamiltonians. Particularly application of the single-impurity Anderson model (SIAM)~\cite{And} was
very successful allowing a quantitative correlation of transport, thermodynamic and magnetic
properties of the materials with spectroscopic data (see, for example, Ref. \onlinecite{GS}). A
shortcoming of this model, however, is its restriction to an isolated $f$ impurity that ignores
the influence of the periodicity of the lattice on the $f$ state. Anisotropies of physical
properties, therefore, are usually only discussed in terms of crystal field effects, while a ${\bf
k}$ dependence of hybridization is not considered~\cite{hybr}. Taking into account the latter
leads to the periodic Anderson model (PAM) for which, however, realistic approaches are presently
still missing~\cite{PAM}.

Experimentally, in particular angle-resolved photoemission (PE) has been used to search for
inconsistencies of SIAM that make use of PAM necessary. In PE spectra, the Ce $4f$ emissions
reveal a characteristic double peak structure consisting of a spin-orbit split Fermi-level ($E_F$)
feature related to the Kondo resonance and a broad peak at about 2~eV binding energy (BE) that
corresponds roughly to the $4f^0$ final state expected for photoionization of a $4f^1$ ground
state. In the light of PAM, BE and intensity variations of these features as a function of
electron wave vector ${\bf k}$ are expected being particularly strong for the Fermi-level peak. In
fact, in a few cases energy dispersions~\cite{CeBi, CeP, CeSbCeBe, CePtb, CeRh3, CeBe} and ${\bf
k}$-dependent intensity variations~\cite{CePt, CeBe, CePtb} were reported and qualitatively
interpreted in the light of PAM. A quantitative description of the phenomena in relation to the VB
structure however is still lacking.

In the present Letter we report on an angle-resolved $4d \rightarrow 4f$ resonant PE study of
CePd$_3$(111). This compound is characterized by a relatively low VB density of states around the
Fermi energy~\cite{Schneider} and band crossings of the Fermi level may only be observed by
photoemission at certain points in ${\bf k}$ space. CePd$_3$(111) represents, therefore, an ideal
system for the search of ${\bf k}$ dependencies of the 4$f$ emission. While BE variations of the
individual spectral $4f$ components are found to be not larger than 30~meV, strong ${\bf
k}$-dependent intensity variations of the Fermi-level feature with respect to the intensity of the
ionization peak are observed. These variations reflect the periodicity of the lattice and reveal
largest intensity close to the $\overline{\Gamma}$ points of the surface Brillouin zones (BZ). The
phenomenon is analyzed in the framework of a simplified PAM where less probable on-site
double-occupation of $4f$ states is neglected in order to achieve ${\bf k}$ conservation upon
interaction. In this case, the intensity of the Fermi-level feature is directly related to the
dispersive properties of the VB states and becomes particularly large at those points of ${\bf k}$
space where the BE of the VB states approaches $E_F$. Taking VB dispersions from LMTO-slab
calculations good agreement between theory and experiment is achieved.

Angle-resolved resonant PE experiments at the $4d \rightarrow 4f$ absorption threshold were
performed with synchrotron radiation provided by the U49/2-PGM-1 undulator beamline of BESSY II. A
hemispherical Thermo-VG CLAM-IV analyzer tuned to an energy resolution of 25~meV (FWHM) and an
angular resolution better than $1^{\circ}$ was used. Epitaxial CePd$_3$ films with thicknesses of
approximately 100~{\AA} were grown on a clean W(110) substrate by thermal deposition of
stoichiometric amounts of Pd and Ce and subsequent annealing~\cite{Schneider}. The resulting LEED
pattern revealed a sharp (2$\times$2) overstructure with respect to the spots of a pure Pd(111)
film as expected for the (111) surface of CePd$_3$ (cubic AuCu$_3$ structure)~\cite{Schneider}.
During vapor deposition as well as for the PE experiments the substrate temperature was held below
90~K. The base pressure in the ultra-high vacuum system was in the upper 10$^{-11}$-mbar range and
raised only shortly to 1$\times 10^{-9}$~mbar during sample preparation. Oxygen and carbon
contaminations were checked monitoring the respective VB signals and found to be negligible.

\begin{figure}[bt]
\includegraphics{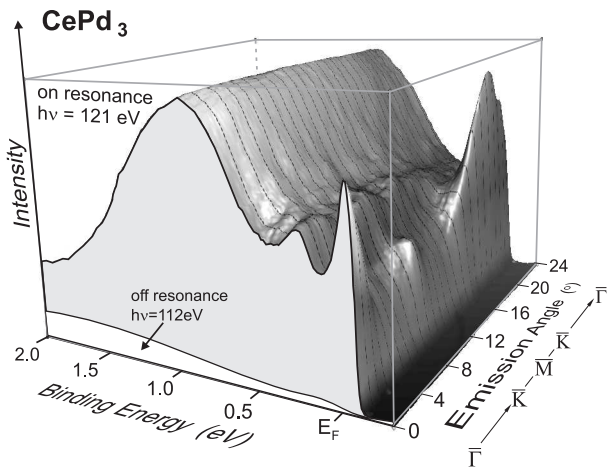}
\caption{Experimental on-resonance PE spectra of CePd$_3$ taken at
the $4d \rightarrow 4f$ excitation threshold ($h\nu=121$~eV) at
different emission angles ($\Theta$) and normalized to maximum
intensity of the ionization peak. The subspectrum illustrates
valence-band contributions as observed off resonance at 112~eV
photon energy, normalized to the same photon flux as the
corresponding on-resonance spectrum ($\Theta =0^{\circ}$).}
\label{fig1}
\end{figure}
Fig.\,1 shows a three dimensional plot of a series of on-resonance
PE spectra taken at 121~eV photon energy and different emission
angles corresponding to a variation of the parallel component of
the ${\bf k}$ vector along the $\overline{\Gamma}-
\overline{\rm{K}}-\overline{\rm{M}}-\overline{\rm{K}}-\overline{\Gamma}$
direction in the surface BZ. At this photon energy the $4f$
photoemission cross section is strongly enhanced by a Fano
resonance, while contributions from the VB emissions are
negligibly small due to a Cooper minimum of the Pd $4d$ cross
section. The latter is illustrated in Fig.\,1 by the blank
subspectrum area, that shows the VB signal in normal emission
geometry observed at the Fano off resonance ($h\nu$=112~eV)
normalized to the same photon flux as the corresponding
on-resonance spectrum. The on-resonance spectra are normalized to
maximum intensity of the ionization peak. The Fermi-level peak
reveals large intensity close to the $\overline{\Gamma}$ point.
Upon leaving the $\overline{\Gamma}$-point region, the intensity
decreases strongly and arrives again at a small local maximum in
the region of the $\overline{\rm{M}}$ point. The spin-orbit split
side-band at about 300~meV BE, on the other hand, reveals almost
constant intensity and increases its BE by about 30~meV when going
from $\overline{\Gamma}$ to $\overline{\rm{M}}$. Thus, the
intensity variation of the Fermi-level peak cannot be ascribed to
photoelectron diffraction effects but reflects an intrinsic
electronic property. Similar, although less dramatic effects have
been reported for CePt$_{2+x}$~\cite{CePt, CePtb} and CeBe$_{13}$
~\cite{CeBe} where the intensity variations were tentatively
ascribed to a dispersing band that crosses the Fermi
energy~\cite{CePt} or a  $\bf k$-vector dependence of
hybridization~\cite{CePtb}.

Starting point of our analysis of the measured PE spectra is the PAM:
\begin{eqnarray}
H &=& \sum_{{\bf k},\sigma} \varepsilon({\bf k})
           d_{{\bf k}\sigma}^{+} d_{{\bf k}\sigma}^{}
     + \sum_{{\bf k},\sigma} \varepsilon_f ({\bf k})
           f_{{\bf k}\sigma}^{+} f_{{\bf k}\sigma}^{}  \nonumber\\
  && + \sum_{{\bf k},\sigma} V_{\bf k}(\varepsilon)
          \left(d_{{\bf k}\sigma}^{+}f_{{\bf k} \sigma}^{}+
               f_{{\bf k} \sigma}^{+}d_{{\bf k}\sigma}^{}\right) \nonumber\\
  && + \frac {U_{ff}}{2} \sum_{i,\sigma}
           n_{i,\sigma}^f n_{i,-\sigma}^f \nonumber\\
  &=& \sum_{{\bf k}} h_0 ({\bf k}) + u ,
\end{eqnarray}
where the extended VB states $|{\bf k}\sigma \rangle$ have a
dispersion $\varepsilon({\bf k})$ and are described by creation
(annihilation) operators $d_{{\bf k}\sigma}^{+}$ ($d_{{\bf
k}\sigma}^{} $). The operator $f_{{\bf k}\sigma}^{+}$ creates a
$f$ electron with momentum ${\bf k}$, spin $\sigma$, and energy
$\varepsilon_f ({\bf k})$. We assume that a non-hybridized $f$
band has no dispersion: $\varepsilon_f ({\bf k})=\varepsilon_f$.
The two electron subsystems (VB and $f$ states) are coupled via a
hybridization $V_{\bf k}(\varepsilon)$, and finally $U_{ff}$ is
the Coulomb repulsion between two $f$ electrons localized on the
same lattice site. The transformation of the latter term to a
${\bf k}$ representation leads to a mixing of states with
different ${\bf k}$ values, which makes the problem difficult to
handle quantitatively. However, if $U_{ff}$ is sufficiently large
with respect to $\vert \varepsilon_f \vert$ and $V_{\bf k}$,
contributions of the $4f^2$ configuration to the ground and
excited states become negligibly small as compared to
contributions of the $4f^0$ and $4f^1$ configurations. In fact,
this condition is roughly fulfilled for Ce transition-metal
compounds. In these systems $U_{ff}$ amounts to 7~eV, while
$\vert \varepsilon_f \vert$ is of the order of 1~eV and the
hybridization parameter is even less than $\vert \varepsilon_f
\vert$~\cite{GS,Hayn}. For $U_{ff} \to \infty$ the probability to
find two $f$ electrons localized on the same site becomes zero,
and the problem could be solved by diagonalizing the hamiltonian
$h_0 ({\bf k})$ that describes a coupling of a $f$ electron with
the energy $\varepsilon_f$ to VB states with a specific wave
vector ${\bf k}$. In this way the problem formally reduces to the
one of SIAM with the only difference that the density of states
(DOS) used in SIAM is now replaced by a $\bf k$-dependent energy
distribution of states~\cite{Hayn}. Since results of SIAM are
usually not strongly affected by the neglecting of 4$f^2$
contributions, one may expect that at least qualitatively our
model leads to a correct description of the experimental data.

As a first step of our theoretical treatment we performed calculations of the VB structure of
CePd$_3$ by means of the LMTO method~\cite{LMTO}. In the calculations Ce was replaced by La in
order to suppress contributions of Ce $4f$ states to the VB DOS. The second step was the
calculation of the spectral function using the simplified PAM as described above. Values taken
directly from the band-structure calculations were the energies of the valence bands
$\varepsilon({\bf k})$ as well as the coefficients $c_l(\varepsilon,{\bf k})$ of the $l$-projected
local expansion of the Bloch functions at the La site. Due to symmetry requirements the localized
$4f$ states couple only to those VB states that reveal non-negligible local $f$ contributions
inside the rare-earth atomic sphere. Consequently, the hybridization matrix element was chosen to
be proportional to the respective coefficients:
\begin{equation}
 V_{\bf k}(\varepsilon)=\Delta \cdot c_f(\varepsilon,{\bf k})
\end{equation}
whereby $\Delta$ (like $\varepsilon_f$) is treated as a constant, but adjustable parameter.

In the kinetic energy range of the experiment ($\sim$\,115~eV) the
mean free path of the photoelectrons amounts to only $\sim$
4.5~\AA ~(Ref. \onlinecite{Hufner}). In order to account for
surface effects in the VB DOS, we calculated the electronic
structure of a five-layer slab constructed from (111) atomic
planes. Then, the PE spectral functions were simulated as
superposition of weighted contributions from different atomic
layers where the third atomic layer (in the middle of the slab)
was considered as bulk. The weights were estimated to be equal to
0.41 for the surface layer, 0.24 for the second layer, and 0.35
for the bulk assuming an exponential dependence of the
photoelectron escape probability.

\begin{figure}[tb]
\includegraphics{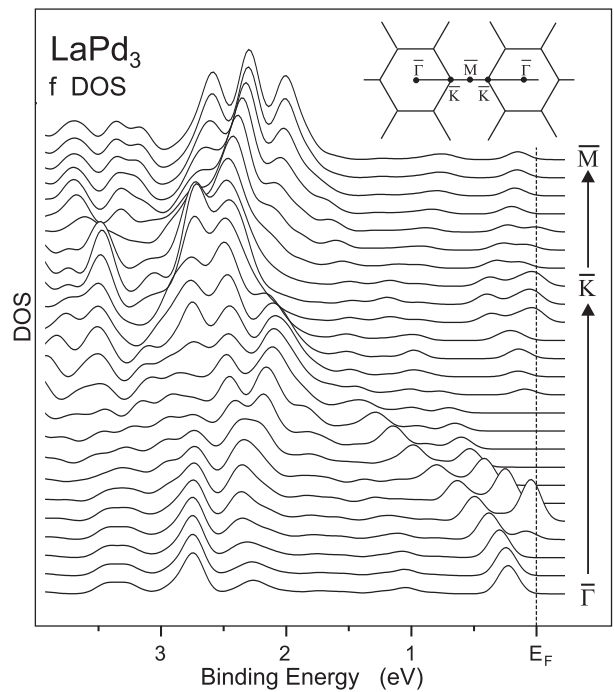}
\caption{Local $f$ DOS ($|c_f(\varepsilon, {\bf k})|^2$ values broadened by a Gaussian of 0.2~eV
FWHM) inside the La atomic sphere of a LaPd$_3$(111) surface layer calculated for different points
along the $\overline{\Gamma}-\overline{\rm{K}}-\overline{\rm{M}}$ direction of the surface BZ. The
results are almost independent from the actual choice of the wave vector component perpendicular
to the surface, since the dispersion of energy bands is found to be negligible small in this
direction.} \label{fig2}
\end{figure}
Fig.\,2 shows the $f$ contributions to the local DOS of
LaPd$_3$(111) along the
$\overline{\Gamma}-\overline{\rm{K}}-\overline{\rm{M}}$ direction
in the surface BZ for the outermost atomic layer. Near the
$\overline{\Gamma}$ point energy bands crossing $E_F$ are mainly
derived from Pd $p$ states with noticeable admixture of La and Pd
$d$ states. At the La site, these bands reveal finite $f$
character. On a half way between $\overline{\Gamma}$ and
$\overline{\rm{K}}$ no bands are found just below $E_F$. In the
vicinity of the BZ borders, however, bands appear again in this
energy region. Near the $\overline{\rm{K}}$ point as well as
between the $\overline{\rm{K}}$ and $\overline{\rm{M}}$ points
weak $f$ contributions to the local DOS are visible. The
respective bands are mainly derived from La $d$ as well as Pd $p$
and $d$ states. Strong peaks below 2~eV BE are related to Pd 4$d$
derived bands that also show finite $f$ character at the La sites.

Calculated 4$f$ spectral functions for the $\overline{\Gamma}$ and
$\overline{\rm{K}}$ points are presented in Fig.\,3 in comparison
with the experimental data.
\begin{figure}[tb]
\includegraphics{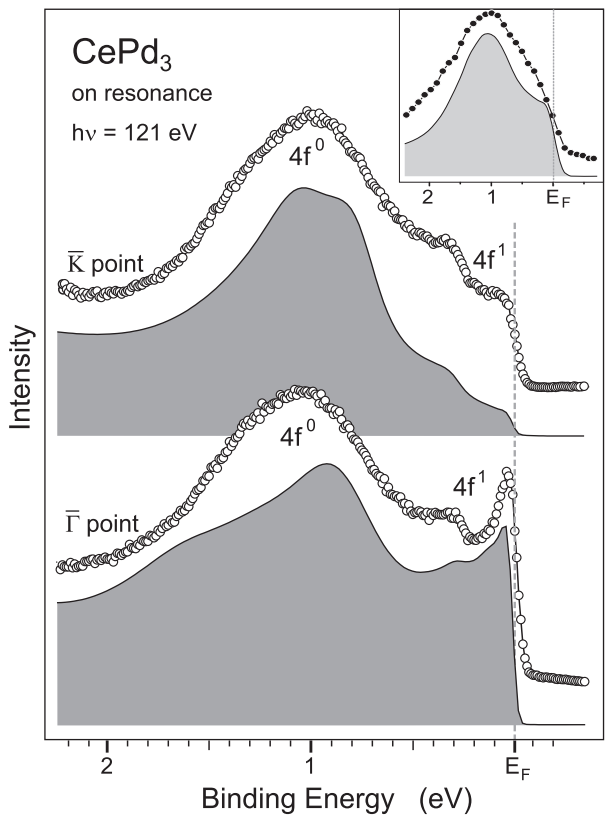}
\caption{Calculated 4$f$ spectral functions (shadowed subspectra: $U_{ff}=\infty$,
$\Delta=0.86$~eV) for the $\overline{\Gamma}$ and $\overline{\rm{K}}$ points in comparison with
the experimental data (open circles). The inset shows the angle-integrated spectrum of CePd$_3$
together with the simulations within SIAM from Ref. 17 where the peak at $E_F$ is not observed,
since it appears only at certain ${\bf k}$ points.} \label{fig3}
\end{figure}
The spectral functions were obtained using similar parameter values as in SIAM~\cite{uff,Hayn}:
$\varepsilon_f =$1.30~eV and $\Delta =$0.86~eV (for the surface atomic layer, $\varepsilon_f$ was
assumed to be shifted by 0.20~eV to higher BE). An energy-dependent life-time broadering parameter
in the form $\Gamma_L(E) = 0.01$ eV $+ 0.12E$ was used, where $E$ denotes the BE with respect to
$E_F$. The calculated spectra were additionally broadened with a Gaussian $(\Gamma_G = 0.03$ eV$)$
to simulate finite instrumental resolution and an integral background was added to take into
account inelastic scattering events. Close to the $\overline{\Gamma}$ point the strong VB $f$
contributions near $E_F$ lead to a large (spin-orbit split) peak at $E_F$ whereas in the spectra
at the $\overline{\rm{K}}$ point only two shoulders are obtained in this energy region due to the
reduced $f$ character of the respective VB states. For ${\bf k}$ points, where the VB $f$
contributions near $E_F$ are negligible, this spectral structure disappears.

As evident from Fig.\,2 there is a Fermi-level crossing of a
second band at about 25\% of the
$\overline{\Gamma}-\overline{\rm{K}}$ distance that should lead to
even larger Fermi-level peak at this $\bf k$ point than around the
$\overline{\Gamma}$ point. Apart from the fact that the respective
angular deviation (1.5$^{\circ}$ at $h\nu$=121 eV) is in the order
of the accuracy of our experiment, a rigid shift of $E_F$ by
0.2~eV to lower or higher BE would dispose this discrepancy.
Energy shifts of this order of magnitude may be caused by
variations of the potential in the rare-earth atomic sphere
caused, e.g. by substitution of Ce by La in the electronic
structure calculations.

In the calculated spectra dispersive substructures appear in the region of the ionization peak are
not observed in the experimental data. The substructures reflect interactions of the ionized $4f$
state with VB states of the same BE and are only weakly related to the ground state properties of
the system~\cite{Kuch}. At least part of the observed discrepancy may be attributed to the
neglecting of the $f$-$f$ correlation term: First, a finite value of $U_{ff}$ would lead to a
certain integration over ${\bf k}$ space and a smearing-out of substructures. The same mechanism
would also cause the appearance of finite Fermi-level features even at ${\bf k}$ points, where no
VB states are found close to $E_F$, in agreement with the experiment. Second, neglecting of the
correlation term leads to an overestimation of $\Delta$,  since contributions to the Fermi-level
peak, that are due to direct $4f^2 \rightarrow 4f^1$ photoemission processes, are attributed to
$4f^0 - 4f^1$ configuration interactions. A smaller value of $\Delta$, however, would immediately
reduce the dispersion of the ionization-peak substructures.

{\it In conclusion}, intensity variations of the Fermi-energy PE peak of CePd$_3$ have been
observed that could be related to a ${\bf k}$ dependence of hybridization in the light of PAM. Our
simple approach has the form of SIAM with a direction-dependent choice of the hybridization
$V_{\bf k}(\varepsilon)$. This suggests that similar approaches may be applied to describe
anisotropies in the thermodynamic, magnetic and transport properties of mixed-valent and
heavy-fermion compounds. Reasonable predictions of the ${\bf k}$ dependence of $V_{\bf
k}(\varepsilon)$ may be obtained from LDA band-structure calculations that like our approach
reveal large hybridization at those points in ${\bf k}$ space where $f$ states are intersected by
VB states. Analyses of angle-integrated PE spectra in the light of SIAM are expected to
underestimate hybridization strength and overestimate $f$ occupation. This is due to the fact that
hybridization is particularly large at the Fermi surface where conduction bands cross the Fermi
energy while angle-integrated experiments probe also other regions of ${\bf k}$ space that are far
away from the Fermi surface and reveal only small hybridization.

\section*{Acknowledgement}

This work was supported by the Deutsche Forschungsgemeinschaft, SFB 463, projects B2, B4, B11, and
B16. Expert assistance by R. Follath and other staff members of BESSY is acknowledged.

%
%
%


\begin{thebibliography}{99}
\bibitem {AAA}
G.R. Steward, Rev. Mod. Phys. {\bf 73}, 797 (2001).


\bibitem {And}
P.W. Anderson, Phys. Rev. {\bf 124}, 41 (1961).


\bibitem {GS}
O. Gunnarsson and K. Sch\"onhammer, Phys. Rev. Lett. {\bf 50}, 604
(1983); Phys. Rev. B {\bf 28}, 4315 (1983); Phys. Rev. B {\bf 31},
4815 (1985).


\bibitem {hybr}
S. Zhang and P.M. Levy, Phys. Rev. B {\bf 40}, 7179 (1989); J.
Kitagawa, N. Takeda, M. Ishikawa, T. Yoshida, A. Ishiguro, N.
Kimura, and T. Komatsubara, Phys. Rev. B {\bf 57}, 7450 (1998).





\bibitem {PAM}
A.N. Tahvildar-Zadeh, M. Jarrell, and J.K. Freericks, Phys. Rev.
Lett. {\bf 80}, 5168 (1998); M.-W. Xiao, Z.-Z. Li, and Wang Xu,
Phys. Rev. B {\bf 65}, 235122 (2002).



\bibitem {CeBe}
A.B.~Andrews, J.J. Joyce, A.J. Arko, Z. Fisk, and P.S.
Riseborough, Phys. Rev. B {\bf 53}, 3317 (1996).



\bibitem {CeBi}
H.~Kumigashira, S.–H. Yang, T. Yokoya, A. Chainani, T. Takahashi,
A. Uesawa, T. Suzuki, O. Sakai, and Y. Kaneta, Phys. Rev. B {\bf
54}, 9341 (1996).



\bibitem {CeP}
H. Kumigashira, S.–H. Yang, T. Yokoya, A. Chainani, T. Takahashi,
A. Uesawa, and T. Suzuki, Phys. Rev. B {\bf 55}, R3355 (1997).


\bibitem {CeSbCeBe}
A.J. Arko, J.J. Joyce, A.B. Andrews, J.D. Thompson, J.L. Smith, D. Mandrus, M.F. Hundley, A.L. 
Cornelius, E. Moshopoulou, Z. Fisk, P.C. Canfield, and A. Menovsky, Phys. Rev. B {\bf 56}, R7041 
(1997).


\bibitem {CePtb}
M. Garnier, D. Purdie, K. Breuer, M. Hengsberger, and Y. Baer,
Phys. Rev. B {\bf 56}, R11399 (1997).


\bibitem {CeRh3}
J. Boysen, P. Segovia, S.L. Molodtsov, W. Schneider, A. Ionov, M.
Richter, and C. Laubschat, J. of Alloys and Compounds {\bf
275-277}, 493 (1998).


\bibitem {CePt}
A.B. Andrews, J.J. Joyce, A.J. Arko, J.D. Thompson, J. Tang, J.M. Lawrence, and J.C. Hemminger, 
Phys. Rev. B {\bf 51}, 3277 (1995).


\bibitem {Schneider}
W. Schneider, S.L. Molodtsov, M. Richter, Th. Gantz, P. Engelmann,
and C. Laubschat, Phys. Rev. B {\bf 57}, 14930 (1998).


\bibitem {Hayn}
R. Hayn, Yu. Kucherenko, J.J. Hinarejos, S.L. Molodtsov, and C.
Laubschat, Phys. Rev. B {\bf 64}, 115106 (2001).


\bibitem {LMTO}
O.K.~Andersen, Phys. Rev. B {\bf 12}, 3060 (1975).


\bibitem {Hufner}
S.~H\"ufner, {\it Photoelectron Spectroscopy} (Springer, Berlin,
1996)

\bibitem {uff}
Yu. Kucherenko, M. Finken, S.L. Molodtsov, M. Heber, J. Boysen, C.
Laubschat, and G. Behr, Phys. Rev. B {\bf 65}, 165119 (2002).


\bibitem {Kuch}
Yu. Kucherenko, M. Finken, S.L. Molodtsov, M. Heber, J. Boysen, G. Behr, and C. Laubschat, Phys. 
Rev. B {\bf 66}, 165438 (2002).



\end{thebibliography}
\end{document}